\documentclass[aps,pra,reprint,superscriptaddress,nofootinbib]{revtex4-1}
\usepackage{graphicx}% Include figure files
\usepackage{bm}% bold math
\usepackage{verbatim}% bold math
\usepackage{overpic}
\usepackage[usenames,dvipsnames]{xcolor}
\usepackage[colorinlistoftodos]{todonotes}
\usepackage{siunitx}
\usepackage{soul}
\usepackage{comment}

\begin{document}
\title{Degenerate states, emergent dynamics and fluid mixing by magnetic rotors}
\author{Takuma Kawai}
\affiliation{Graduate School of Engineering Science, Osaka University, Toyonaka 5608531, Japan}
\author{Daiki Matsunaga}
\email{daiki.matsunaga@me.es.osaka-u.ac.jp}
\affiliation{Division of Bioengineering, Graduate School of Engineering Science, Osaka University, Toyonaka 5608531, Japan}
\affiliation{Rudolf Peierls Centre for Theoretical Physics, University of Oxford, Oxford OX1 3PU, UK}
\author{Fanlong Meng}
\email{fanlong.meng@itp.ac.cn}
\affiliation{CAS Key Laboratory for Theoretical Physics, Institute of Theoretical Physics, Chinese Academy of Sciences, Beijing 100190, China}
\affiliation{Max Planck Institute for Dynamics and Self-Organization (MPIDS), G\"ottingen 37077, Germany.}
\affiliation{Rudolf Peierls Centre for Theoretical Physics, University of Oxford, Oxford OX1 3PU, UK}
\author{Julia M. Yeomans}
\affiliation{Rudolf Peierls Centre for Theoretical Physics, University of Oxford, Oxford OX1 3PU, UK}
\author{Ramin Golestanian}
\affiliation{Max Planck Institute for Dynamics and Self-Organization (MPIDS), G\"ottingen 37077, Germany.}
\affiliation{Rudolf Peierls Centre for Theoretical Physics, University of Oxford, Oxford OX1 3PU, UK}

\date{\today}

\begin{abstract}
We investigate the collective motion of magnetic rotors suspended in a viscous fluid under an uniform rotating magnetic field. The rotors are positioned on a square lattice, and low Reynolds hydrodynamics is assumed. For a $3 \times 3$ array of magnets, we observe three characteristic dynamical patterns as the external field strength is varied: a synchronized pattern, an oscillating pattern, and a chessboard pattern. The relative stability of these depends on the competition between the  energy due to the external magnetic field and the energy of the magnetic dipole-dipole interactions among the rotors. We argue that  the chessboard pattern can be understood as an alternation in the stability of two degenerate states, characterized by striped and spin-ice configurations, as the applied  magnetic field rotates. For larger arrays, we observe propagation of slip waves that are similar to metachronal waves. The rotor arrays have potential as microfluidic devices that can mix fluids and create  vortices of different sizes.
\end{abstract}

% insert suggested PACS numbers in braces on next line
% insert suggested keywords - APS authors don't need to do this
%\keywords{}

%\maketitle must follow title, authors, abstract, \pacs, and \keywordshttps://ja.overleaf.com/project/5d4b7199d90a9e6d4df45d50
\maketitle

\section{Introduction}
Magnetic forces provide a useful way of driving small units in viscous fluids. For example, in recent years magnetic driving has been extensively used in microfluidic devices, for applications such as magnetic swimmers \cite{Ogrin2008,Zhang2009,Meng2018,Hamilton2018}, magnetic pumps \cite{Leach2006,Matsunaga2019} and cilia \cite{Coq2011,Wang2013,Tsumori2016,Hanasoge2018,Meng2019}, and to facilitate particle sorting and segregation \cite{Matsunaga2017,Matsunaga2018,Zhou2017,Zhang2019}. Many of the approaches are designed to control a single magnetic unit or to actuate many units in exactly the same manner. For further development of magnetically-actuated devices, it would be desirable to be able to control the dynamics of the magnetic units just by changing the external magnetic field. However, it is not immediately obvious how to achieve different behavior among different magnetic units when they are driven by the same external magnetic driving mechanism.

In a recent work \cite{Matsunaga2019}, we analyzed the collective motion of a square array of magnetic rotors and found a surprisingly rich dynamics under an external magnetic field which oscillated along one axis of the array. By changing the relative strength of the external field and the dipolar interactions between the rotors, different collective rotational patterns emerged. When the dipole interaction was dominant the rotors swung back and forth, clockwise or counterclockwise  in alternating stripes. When the external field dominated over the dipolar interactions, the rotors underwent full rotations with different quadrants of the array turning in different directions.

\begin{figure}[b]
  \begin{center}
   \includegraphics[width=0.99\hsize]{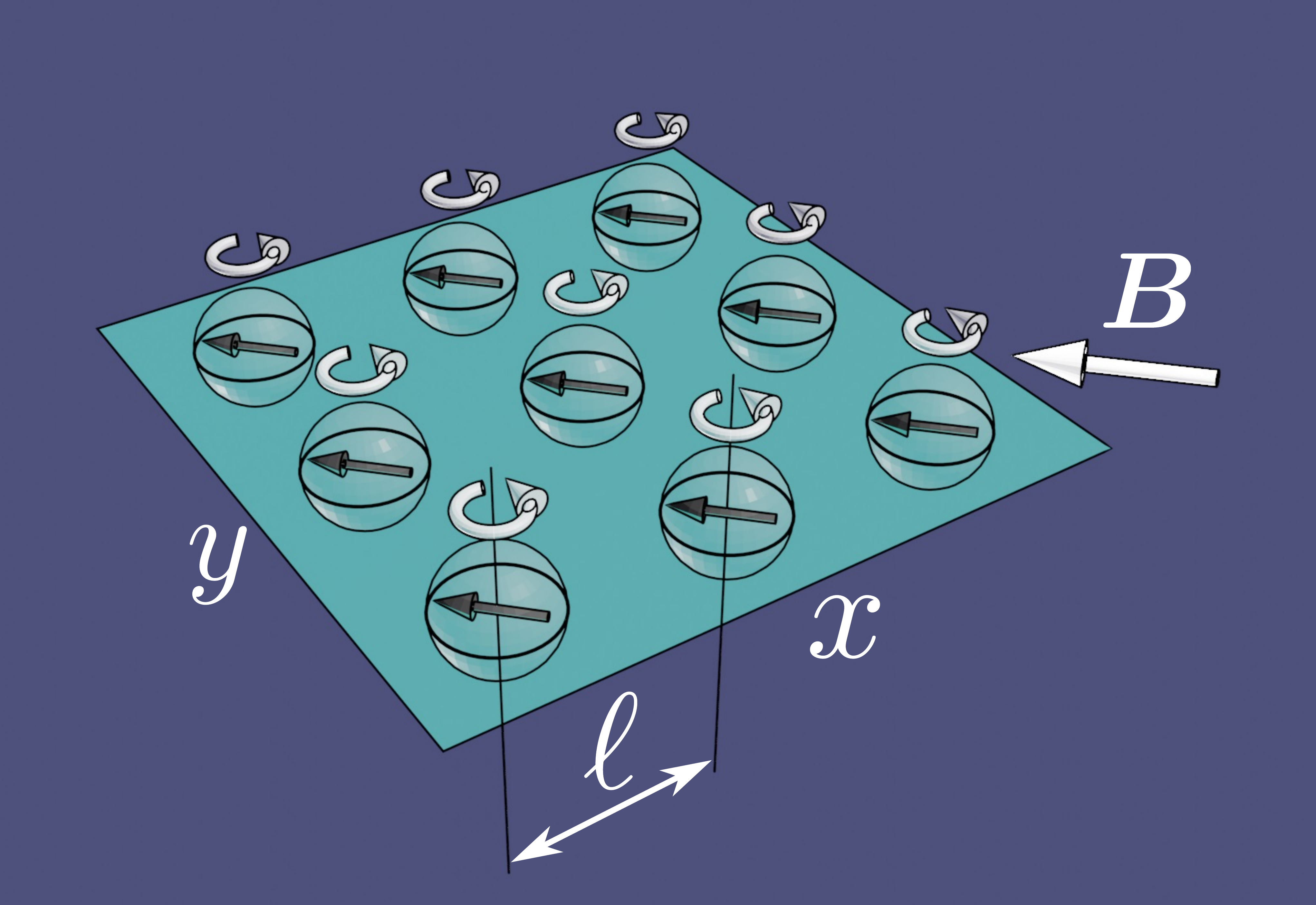}
  \end{center}
    \caption{Schematic representation of the array of magnets. The rotors are positioned on a square grid in the $xy$-plane with spacing $\ell$. Arrows in the rotors represent magnetic dipole moments. The magnets rotate about the $z$-axis driven by an external field ${\bm B}$.}
\label{fig:setup}
\end{figure}

In the present paper, we extend these results to analyze the motion of the rotor array under a rotating magnetic field (see Fig.~\ref{fig:setup} for a schematic representation of our system). This enables us to identify three different collective rotational patterns as the balance of the torque due to the external field and the torque due to dipolar interactions is varied. In the first regime that occurs for strong fields, we observe synchronized patterns in which all the magnetic units rotate in the same direction as the external field. We observe oscillating patterns for weak field strengths, when the rotors cannot achieve net rotation. We also observe a most interesting chessboard pattern, which appears when the contributions of the two torques are comparable. In this pattern, the rotors rotate in a direction opposite to their closest neighbouring rotors.

We examine system sizes up to $100 \times 100$ rotors, and report the observation of the propagation of novel slip waves that are similar to the metachronal waves that arise from coordinated motion of cilia \cite{Uchida2010,Osterman2011,Golestanian2011,Brumley2012,Elgeti2013,Narematsu2015,Brumley2015}.
We demonstrate that our system can be used as a microfluidic device for mixing and pumping, as it can drive flow fields with different vortex sizes and mixing length scales.

\section{Description of the system and governing equations}
\subsection{The system}
Our system consists of magnetic rotors positioned on a square lattice in $xy$-plane with a lattice spacing $\ell$ as shown in Fig.~\ref{fig:setup}. The total number of rotors is $N = N_x N_y$, where $N_x$ and $N_y$ are the number of rotors in the $x$- and $y$-directions, respectively. The rotors are embedded in an unbounded fluid domain with viscosity $\eta$ and density $\rho$. The magnetic dipoles have no translational degrees of freedom because they are fixed in space, but each has a single rotational degree of freedom about the $z$-axis. Every rotor is taken to be a sphere with radius $a$, which has a magnetic dipole moment
\begin{equation}
    \bm{m}_i = (m \cos \phi_i, m \sin \phi_i, 0),
\end{equation}
where $m$ is the magnitude and $\phi_i$ is the polar angle characterizing the direction.

The rotors are driven by a rotating magnetic field
\begin{equation}
    \bm{B}(t) = (B \cos \theta, B \sin \theta, 0)
\end{equation}
where $B$ is the field strength, $t$ is the time, and $\theta (t) = 2\pi ft$ is the polar angle of the external field that is rotating at a frequency $f$. We assume that hydrodynamic is in the viscous regime (low Reynolds number ${\rm Re}= a^2 \rho f/\eta \ll 1$).

\subsection{Governing equations}
Each rotor experiences a magnetic torque that comprises contributions from the external field $T^{\rm ext}$, and the dipole-dipole interactions $T^{\rm dd}$. The torque contributions on the $i$-th rotor are
\begin{eqnarray}
  T_i^{\rm ext}
  &=& (\bm{m}_i \times \bm{B})\cdot \hat{\bm{e}}_z = m B \sin(\theta - \phi_i),
  \label{eq:Text} \\
  T_i^{\rm dd}
  &=& \left(\bm{m}_i \times \frac{\mu_0}{4 \pi} \sum_{j \neq i}^N \frac{3 (\bm{m}_j \cdot \bm{n}_{ij}) \bm{n}_{ij} - \bm{m}_j}{r_{ij}^3}\right) \cdot \hat{\bm{e}}_z, \nonumber \\
  &=& - \frac{\mu_0 m^2}{4\pi} \sum_{j \neq i}^N \frac{3\sin(\phi_i+\phi_j-2\gamma_{ij})+\sin(\phi_i-\phi_j)}{2r_{ij}^3} \nonumber \\
    \label{eq:Tdd}
\end{eqnarray}
where $\bm{r}_i$ is the position of $i$-th rotor, $\bm{r}_{ij} = \bm{r}_j - \bm{r}_i$, $r_{ij} = |\bm{r}_{ij}|$, $\bm{n}_{ij} = (\cos \gamma_{ij}, \sin \gamma_{ij}, 0) = (\bm{r}_j - \bm{r}_i)/r_{ij}$, and $\hat{\bm{e}}_z$ is the unit vector along the $z$ direction. Torque balance in the viscous regime gives the angular velocity $\omega_i$ of a rotor as
\begin{equation}
  \omega_i = \frac{d \phi_i}{dt} = \frac{T_i}{8 \pi \eta a^3},\label{eq:omega1}
\end{equation}
where $T_i = T_i^{\rm ext}+ T_i^{\rm dd}$ and $8\pi \eta a^3$ is the friction constant for the rotation of a sphere \cite{Kim1991}. Note that we have ignored the hydrodynamic interaction between rotors by assuming that the rotor radius $a$ is sufficiently small compared to the lattice distance $a \ll \ell$. The leading order effect of hydrodynamic coupling is discussed in Appendix \ref{app:fluid}.

The potential energies due to the external magnetic field, $U^{\rm ext}$, and magnetic interaction, $U^{\rm dd}$, are
\begin{eqnarray}
  U^{\rm ext}&=& - \sum_{i}^N \bm{m}_i \cdot \bm{B} =  -m B \sum_{i}^N \cos (\theta - \phi_i), \label{eq:Uext} \\
  U^{\rm dd} &=& -\frac{\mu_0 m^2}{4\pi}\sum_{i=1}^{N} \sum_{j=1}^{i-1} \frac{3\cos(\phi_i+\phi_j-2\gamma_{ij})+\cos(\phi_i-\phi_j)}{2r_{ij}^3}. \nonumber \\ \label{eq:Udd}
\end{eqnarray}

The velocity field and the vorticity field at a position $\bm{x}$ are described by the rotlet \cite{Kim1991}
\begin{eqnarray}
  \bm{v}\left(\bm{x}\right)
  &=& \frac{1}{8\pi \eta}\sum_i^N \left\{ \frac{1}{\left| \bm{x}-\bm{r}_i \right|^3} \bm{T}_i \times \left( \bm{x}-\bm{r}_i \right) \right\}, \label{eq:v}\\
  \bm{\Omega}\left(\bm{x}\right) &=& \nabla \times \bm{v}\left(\bm{x}\right). \label{eq:omega}
\end{eqnarray}
When the voriticity observation point is in the $z = 0$ plane, it can be simply rewritten as
\begin{equation}
  \Omega_z = -\frac{1}{8\pi\eta}\sum_i^N\frac{T_i}{\left|\bm{x}-\bm{r}_i\right|^3}. \label{eq:omega_z}
\end{equation}

\subsection{Dimensionless parameters}
We introduce two dimensionless parameters:
\begin{eqnarray}
  {\rm Mn}  &\equiv& \frac{32\pi^2 \eta a^3 \ell^3 f}{\mu_0 m^2}, \\
  \tau_r f &=& \frac{8 \pi \eta a^3 f}{m B},
\end{eqnarray}
where $\tau_r$ is the characteristic relaxation time of the magnetic dipoles. The Mason number ${\rm Mn}$ characterizes the relative strength of viscous stresses as compared to the magnetic ones \cite{Driscoll2017}, and $\tau_r f$ characterizes the relaxation time of the system as compared to the period of the rotation of the external field.

Equation (\ref{eq:omega1}) can then be rewritten in non-dimensional form:
\begin{equation}
  \omega_i^*=\frac{\omega_i}{f} = \frac{1}{\rm Mn} \,T_i^{{\rm dd}*}+\frac{1}{\tau_r f} \,T_i^{{\rm ext}*}, \label{eq:omegaast}
\end{equation}
where $r_{ij}^* = r_{ij}/\ell$, $T_i^{{\rm ext}*} = T_i^{\rm ext}/(mB)$, and $T_i^{{\rm dd}*} = T_i^{\rm dd} \times 4\pi \ell^3/(\mu_0 m^2)$.
Similarly, we define the dimensionless potential energies as $U^{\rm ext*} = U^{\rm ext} /(mB)$ and $U^{\rm dd*}=U^{\rm dd}\times 4\pi \ell^3 /(\mu_0 m^2)$.

\begin{figure*}
 \includegraphics[width=0.9\hsize]{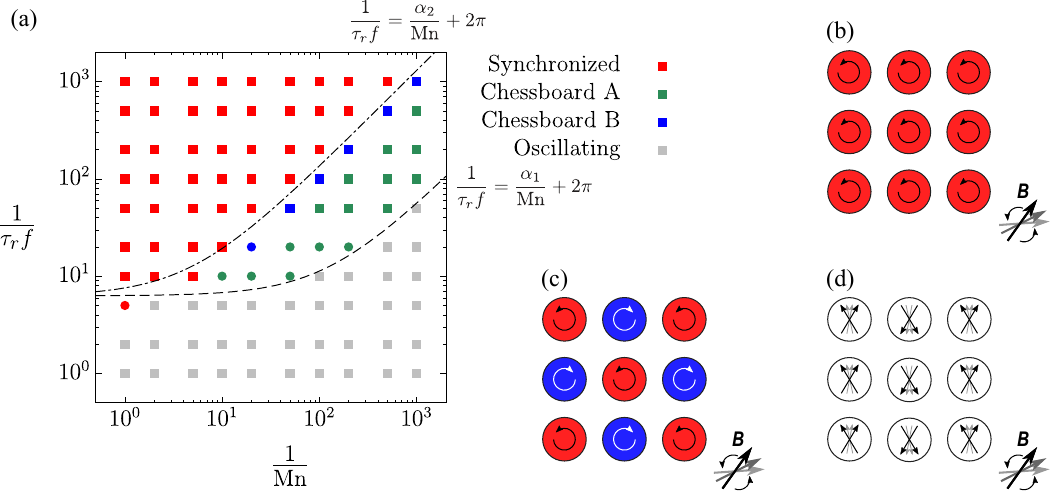}
 \caption{Rotational patterns for the $3\times3$ array of rotors. (a) Phase diagram in the space of the two parameters $1/{\rm Mn}$ and $1/(\tau_r f)$. Square markers represent $|R|=0$ or $1$ and circle markers represent $0<|R|<1$. The difference between chessboards A and B is in the transient orientation of the rotors as shown later in Fig.~\ref{fig:chess-mechanism}(b) and \textbf{Movie 2}. (b) Synchronized pattern: all rotors rotate in the same direction as the external field.  (c) Chessboard pattern: all rotors rotate in the opposite direction to their nearest neighbours. The colors represent the rotational direction: clockwise (blue) and counterclockwise (red). (d) Oscillating pattern: the rotors oscillate instead of performing full rotations. The arrows on the rotors represent the magnetic moment of the rotors.} \label{fig:3x3}
\end{figure*}

\subsection{Numerical method}
We discretize Eq.~(\ref{eq:omegaast}) and follow the time evolution of the rotor angles using a first-order Euler method with a time step $f \Delta t = 10^{-4}$, and the initial orientation angles $\phi_i$ are set randomly. Before the main simulation process starts, a strong external field corresponding to $1/(\tau_r f) = 10^4$ is applied in the $+x$-direction for a dimensionless time $f t = 1$ to align the rotors. When we analyze the average motion of the rotors or the flow field, we run the simulation for 50 cycles and take the average of the last 30 cycles.

\section{Results}

\subsection{Motion of a single rotor}
For completeness, we first summarize the results for a single rotor.% system $N_x = N_y = 1$.
In the absence of dipole interaction, the rotational velocity is determined purely by the torque from the external field $T^{\rm ext}$. The rotational velocity is therefore simply 
\begin{equation}
  \omega(t) = \frac{m B}{8 \pi \eta a^3} \sin (\theta(t) - \phi (t))
\end{equation}
and the time evolution of the phase difference $\Delta = \theta - \phi$ is
\begin{equation}
  \frac{\dot{\Delta}(t)}{f} = 2 \pi - \frac{1}{\tau_r f} \sin \Delta(t).
\end{equation}
For $1/(\tau_r f) > 2 \pi$ the stable solution has $\dot{\Delta} = 0$ and the rotor follows the external field with a constant angular velocity $\omega = 2 \pi f$. For  $1/(\tau_r f) < 2 \pi$, however, there is no solution with
 $\dot{\Delta} \neq 0$: the rotor does not reach a constant velocity but periodically slips to move backwards as it fails to keep up with the external field \cite{Zhang2009,Peyer2013,Meng2019}.
 
\begin{figure*}
\begin{center}
\includegraphics[width = 0.9\hsize]{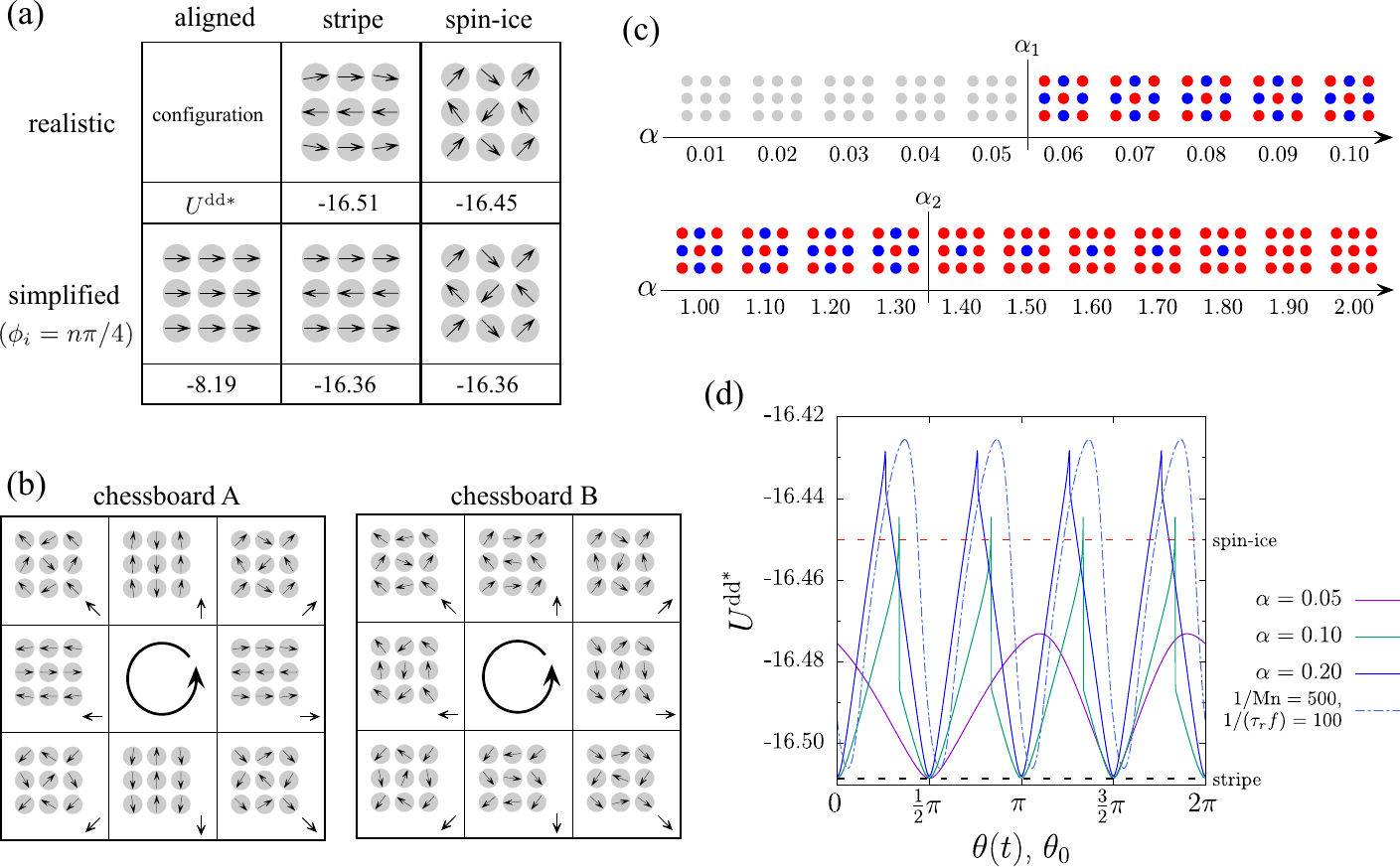}
\caption{The mechanism behind the chessboard rotational pattern.
(a) Stable configurations for $\alpha = 0$ for the $3 \times 3$ magnet array, and the corresponding energies for magnetic dipole interactions $U^{\rm dd*}$.
(b) Stable configurations for different external field angles $\theta$, which are shown by the arrows at the bottom right of each sub-figure. The dimensionless parameters $\alpha=0.5$ and $1.0$ are used to obtain the chessboard A and B patterns respectively.
(c) Rotational directions of the rotors under simulation conditions $1/{\rm Mn} \gg 1$ and $1/{\tau_r f} \gg 1$. The red rotors rotate in the same direction as the external field, whereas the blue rotors rotate in the opposite direction. The gray rotors have no net rotation.
	%These analyses are for deciding boundary parameters between oscillate and chessboard ($\alpha_1$) or chessboard and synchronized ($\alpha_2$).
(d) $U^{\rm dd*}$ as a function of $\theta$ for different simulation conditions. A static magnetic field is used to obtain the lines that have varying $\alpha$ parameters, while a time-varying magnetic field is used for the condition $1/{\rm Mn} = 500$ and $1/(\tau_r f) = 100$.
	}
    \label{fig:chess-mechanism}
  \end{center}
\end{figure*}

\subsection{$3 \times 3$ array}
Figure~\ref{fig:3x3}(a) is a phase diagram showing the dynamical configurations of $3 \times 3$ arrays of magnetic rotors (see also \textbf{Movie 1}). The competition between the two torques $T^{\rm ext}$ and $T^{\rm dd}$, which are controlled by the two dimensionless parameters $1/(\tau_r f)$ and $1/{\rm Mn}$, respectively, determines the rotational patterns. We identify three different responses to the field: ``synchronized" (Fig.~\ref{fig:3x3}(b)), ``chessboard" (Fig.~\ref{fig:3x3}(c)), and ``oscillating" (Fig.~\ref{fig:3x3}(d)) patterns. Note that we introduce a parameter $R$ to characterize the rotational pattern \cite{Matsunaga2019}, defined as
\begin{equation}
R = \frac{\Delta \phi_i}{\Delta \theta},\label{R}
\end{equation}
where $\Delta \phi_i$ and $\Delta \theta$ are the total rotational angles of the rotors and the external field, respectively.

The synchronized pattern appears when the torque due to the external field is dominant over the dipolar interactions $1/(\tau_r f) > 1/{\rm Mn}$, which corresponds to the top left corner of Fig.~\ref{fig:3x3}(a). In this regime, all rotors rotate in the direction of the external field as shown in Fig.~\ref{fig:3x3}(b). The oscillating pattern appears when the dipolar interaction is dominant $1/{\rm Mn} \gg 1/(\tau_r f)$, or when the rotors cannot catch up with the external field $1/(\tau_r f) < 2\pi$ as also seen in the motion of a single rotor. The rotors simply oscillate and cannot achieve a net rotation  because the external field is not strong enough to destroy the configurations favoured by the magnetic interaction, such as the spin-ice pattern \cite{Bramwell2001,Castelnovo2008,Matsunaga2019}. 

The chessboard pattern appears when the contributions of the two torques are comparable, namely, $1/{\rm Mn} \sim 1/(\tau_r f)$. In this regime, all rotors rotate in a direction opposite to their closest neighbours. This is counter-intuitive because half of the rotors are rotating in the opposite direction to the magnetic field. There have been reports of such chessboard-like behaviour in other active matter systems, such as bacterial suspensions \cite{Wioland2016,Thampi2016,Nishiguchi2018}, although they do not operate under the effect of an external driving mechanism unlike our system.

\subsubsection*{The mechanism behind the chessboard pattern}
In order to further understand the rotational patterns, we next apply a static magnetic field, $\bm{B} (\theta_0) = (B \cos \theta_0, B \sin \theta_0, 0)$ where $0 \leq \theta_0 < 2\pi$, and analyze the equilibrium orientation configuration for each angle $\theta_0$. Since the rotors are in the equilibrium state, there is only one relevant dimensionless parameter in the system, namely
\begin{equation}
  \alpha = \frac{\rm Mn}{\tau_r f} = \frac{4 \pi B \ell^3}{\mu_0 m}, \label{eq:alpha}
\end{equation}
which compares the strength of the external field to the typical dipole field \cite{Matsunaga2019}.

When there is no applied external field (i.e. $\alpha = 0$) there exist several equilibrium configurations for a square rotor array such as the stripe and spin-ice configurations shown in Fig.~\ref{fig:chess-mechanism}(a). In a $3 \times 3$ array, the magnetic dipole-dipole interaction energy $U^{\rm dd*}$ is in fact the same in these two configurations for regular alignment of the magnetic moments, i.e., $\phi_i = n\pi/4$ with $n$ as integers. Note, however, that the directions of the magnetic dipoles in finite systems can have minor deviations from the regular alignment defined by $\phi_i = n\pi/4$, which corresponds to infinite systems. This leads to the stripe configuration having slightly lower dipolar potential energy ($U^{\rm dd*} \approx -16.51$) than the spin-ice configuration ($U^{\rm dd*} \approx -16.45$) in the 
$3 \times 3$ array.

Starting from random rotor orientations without an external field, namely $\alpha = 0$, the simulation results in one of these equilibrium configurations depend on the details of the initial angles. If we now impose a static external magnetic field that is comparable to the dipolar field, namely $\alpha \sim 1$, the system changes the pattern to that dictated by  the external field direction $\theta$. As shown in the chessboard A pattern in Fig.~\ref{fig:chess-mechanism}(b), the system prefers stripe-like configurations for external field directions $\theta_0 = n\pi/2$ where $n$ is an integer, while it prefers spin-ice-like configurations for $\theta_0 = (2n + 1)\pi/4$ (see Appendix \ref{app:B}).
Since the chessboard rotational pattern can be obtained by following the sequence of 8 configurations in Fig.~\ref{fig:chess-mechanism}(b), we can understand that it is a consequence of alternate switching between stripe and spin-ice ordering.

Note that we classified chessboard patterns into two different patterns, chessboard A (that corresponds to smaller $\alpha$) and chessboard B (that corresponds to larger $\alpha$), depending on the transient configurations in Fig.~\ref{fig:chess-mechanism}(b). In the chessboard B pattern, configurations at $\theta_0 = n\pi/2$ are not stripe-like patterns (such as chessboard A), as shown in \textbf{Movie 2}.

The chessboard pattern can be seen only for $\alpha \sim 1$. Figure~\ref{fig:chess-mechanism}(d) shows the potential of dipolar interactions $U^{\rm dd*}$. When the external field is weak ($\alpha \ll 1$), as for $\alpha = 0.05$ in Fig.~\ref{fig:chess-mechanism}(d), the system cannot transit from the stripe pattern to the spin-ice pattern because the external field strength is not sufficiently strong for the rotors to jump across the energy barrier $\Delta U^{\rm dd*}$ between the two configurations. As the result, they stay on a single stripe pattern and just result in the oscillating pattern.
When the external field is moderately strong, say for $\alpha = 0.2$, the system can make the transition into all 8 patterns as shown in Fig.~\ref{fig:chess-mechanism} (b) and (d). When the external field is strong ($\alpha \gg 1$) on the other hand, the chessboard pattern cannot be seen because the rotors all align to the external field direction and the stripe or the vortex configurations disappear.

\begin{figure*}[tb]
  \begin{center}
   \includegraphics[width=0.95\hsize]{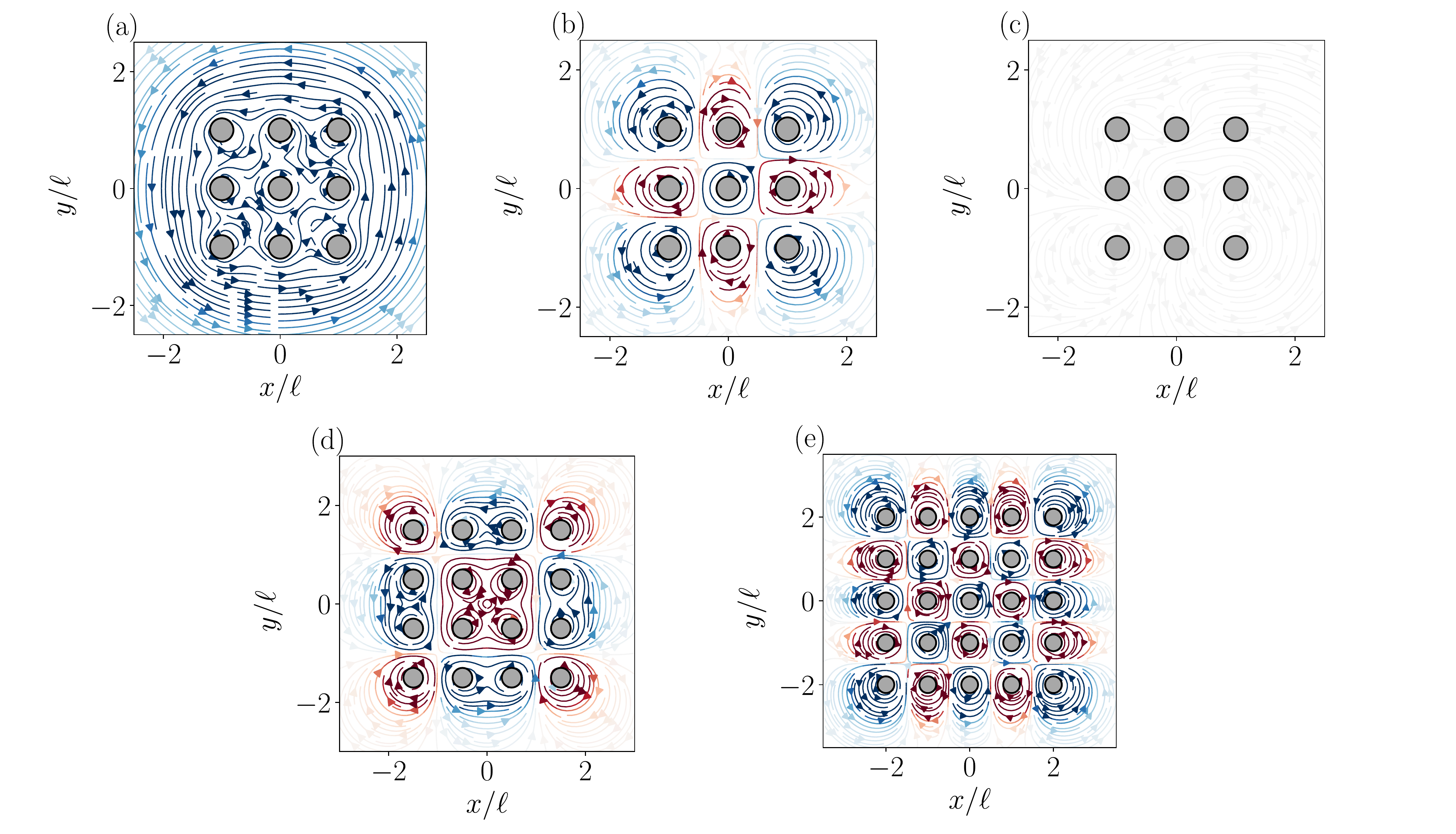}
  \end{center}
 \caption{Streamlines and vorticity patterns for different conditions: (a) synchronized pattern $1/\rm{Mn} = 100$, $1/\tau_r f = 1000$, (b) chessboard pattern ($1/\rm{Mn} = 100$, $1/\tau_r f=100$), (c) oscillating pattern ($1/\rm{Mn} = 100$, $1/\tau_r f = 10$), (d) chessboard pattern for a $4 \times 4$ array ($1/\rm{Mn} = 100$, $1/\tau_r f=100$) and (e) chessboard pattern for a $5 \times 5$ array ($1/\rm{Mn} = 100$, $1/\tau_r f=100$). Colours represent the direction of local vortices: red shows $+z$-rotation while blue shows $-z$-rotation.}
 \label{fig:streamline}
\end{figure*}

\subsubsection*{Predicting the phase boundaries}
Considering a simpler problem allows us to obtain an estimate for the boundaries between the different patterns.
Figure~\ref{fig:chess-mechanism}(c) shows the rotational patterns for different values of $\alpha$. Under a negligible viscous friction, namely $1/{\rm Mn} \gg 1$ and $1/(\tau_r f) \gg 1$, we expect each rotor angle $\phi_i$ to follow the local magnetic field instantaneously. The simplified system shows all three patterns as $\alpha$ is varied. The numerical value for the threshold between the oscillating and the chessboard pattern is $\alpha_1 = 0.05-0.06$ while the corresponding value for the transition between the chessboard and synchronized patterns is $\alpha_2 = 1.3-1.4$. By recalling the definition of the parameter $\alpha$ in Eq.~(\ref{eq:alpha}), the phase boundaries between the three rotational patterns can be estimated using
\begin{equation}
   \frac{1}{\tau_r f} = \frac{\alpha}{\rm Mn} + 2 \pi. \label{eq:boundary}
\end{equation}
The second term of right hand side, $2\pi$, appears because there is a threshold for slipping, namely $1/(\tau_r f) = 2 \pi$, as shown in the single rotor analysis $1/{\rm Mn} \ll 1$. The lines in Fig.~\ref{fig:3x3}(a) correspond to Eq.~(\ref{eq:boundary}) with the parameters $\alpha_1$ and $\alpha_2$. The predictions agree well with the simulation results.

Analytical estimates for the values of $\alpha_1$ and $\alpha_2$ can be obtained by considering the energies of the different patterns. For simplicity, we consider only rotor orientations that satisfy $\phi_i = n \pi/4$ where $n$ is an integer, thus approximating the stripe and spin-ice configurations as shown in Fig.~\ref{fig:chess-mechanism}(a). The corresponding dipole interaction energy $U^{\rm dd*}$ is also shown in the figure. For a given orientation patterns $\phi_i$, the total dimensionless energy can be estimated as
\begin{equation}
U^{\rm tot*}(\phi_i, \theta_0, \alpha) = U^{\rm dd*} (\phi_i) + \alpha U^{\rm ext*} (\phi_i, \theta_0). \label{eq:total_energy}
\end{equation}

An estimate of $\alpha_1$ is obtained by comparing the total energy of the stripe and spin-ice configurations for an external field direction $\theta_0 = \pi/4$. This is based on the observation that the chessboard pattern appears when the total energy of spin-ice configurations is lower than that of stripe configurations. Using $U^{\rm ext*}(\phi_i^{\rm stripe}, \pi/4) = -3\sqrt{2}/2$ and  $U^{\rm ext*} (\phi_i^{\rm ice}, \pi/4) =-3$, and assuming that the energy varies monotonically as a function of $\alpha$, we obtain $\alpha_1 = 0.068$, which is in good agreement with the simulation result.

For $\alpha_2$, the estimate follows from comparing the total energy of the striped and aligned (Fig.~\ref{fig:chess-mechanism}(a)) configurations, for an external field at $\theta_0 = 0$ noting that the synchronized pattern appears when the total energy of aligned configurations is lower than that of the striped ones. Using $U^{\rm ext*}(\phi_i^{\rm stripe}, 0) = -3$ and  $U^{\rm ext*} (\phi_i^{\rm aligned}, 0) =-9$, we obtain $\alpha_2 = 1.36$, which is again in good agreement with the numerical value obtained from our simulations.

\subsubsection*{Velocity field}
The collective rotational patterns can be used to produce complex rotational flow fields \cite{Matsunaga2019}.
Figure~\ref{fig:streamline} shows time-averaged streamlines and vorticity patterns, evaluated by using equations (\ref{eq:v}) and (\ref{eq:omega_z}), respectively. For the synchronized pattern, the flow field is a single large vortex as shown in Fig.~\ref{fig:streamline}(a). Both the rotors and the large vortex rotate in the counterclockwise direction.
Note that the local vorticity direction is opposite to that of the torque, as shown in Eq.~(\ref{eq:omega_z}). 
For the chessboard pattern, each rotor creates a smaller, individual vortex as shown in Fig.~\ref{fig:streamline}(b).
There is no flow between the rotors because the flow field generated by neighbours cancels out.
In the oscillating pattern there is no time-averaged flow field because the net rotation angle of each rotor is zero.

It is well known that mixing at small Reynolds numbers is challenging \cite{Nguyen2004,Lee2011,Aref2017}, because of the absence of inertia. This device would allow the vortex size and the mixing length scale to be controlled by the external magnetic field. Other types of flow fields are also possible for oscillatory magnetic driving mechanisms \cite{Matsunaga2019}.

\begin{figure*}
 \begin{minipage}{0.42\hsize}
   \begin{center}
   \includegraphics[width=\columnwidth]{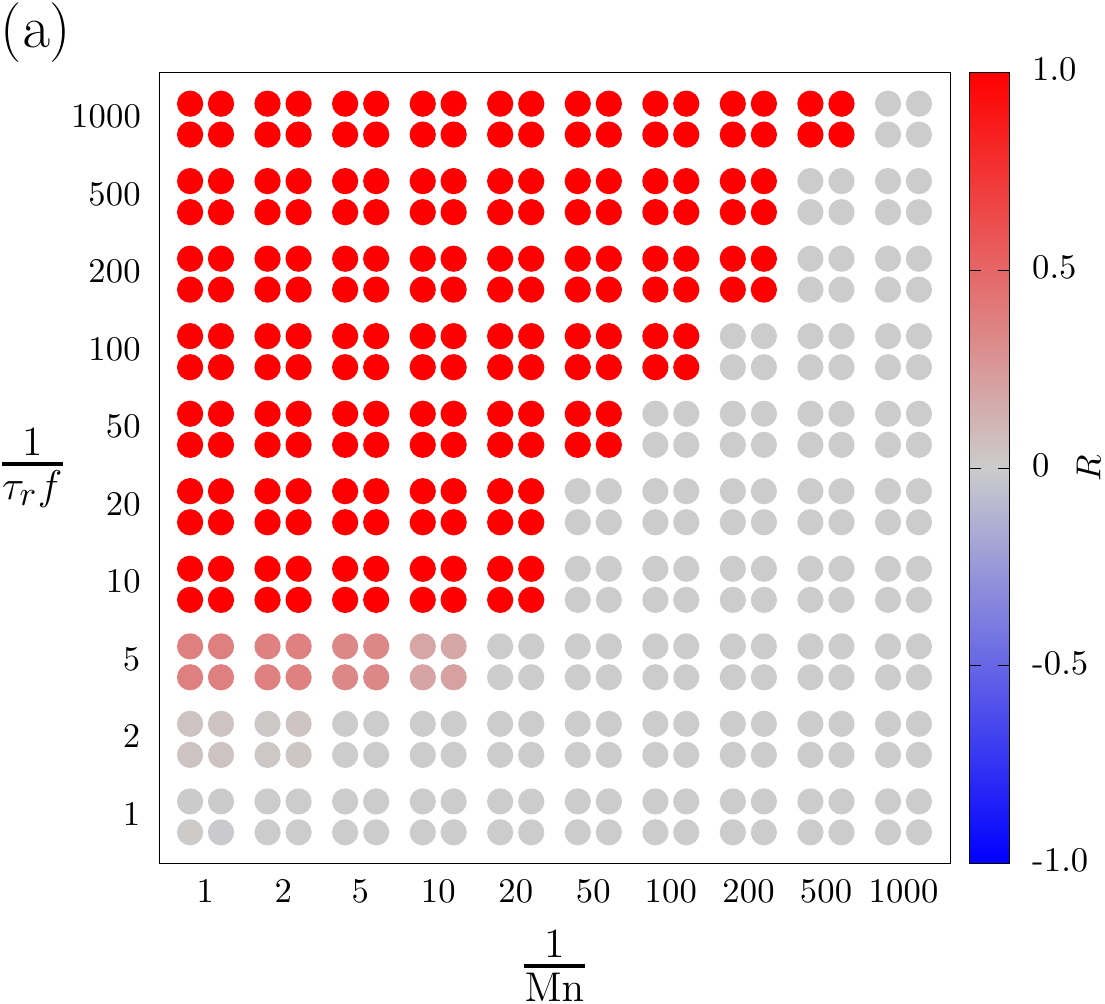}
   \end{center}
 \end{minipage}
   \hspace{0.05\hsize}
 \begin{minipage}{0.42\hsize}
   \begin{center}
   \includegraphics[width=\columnwidth]{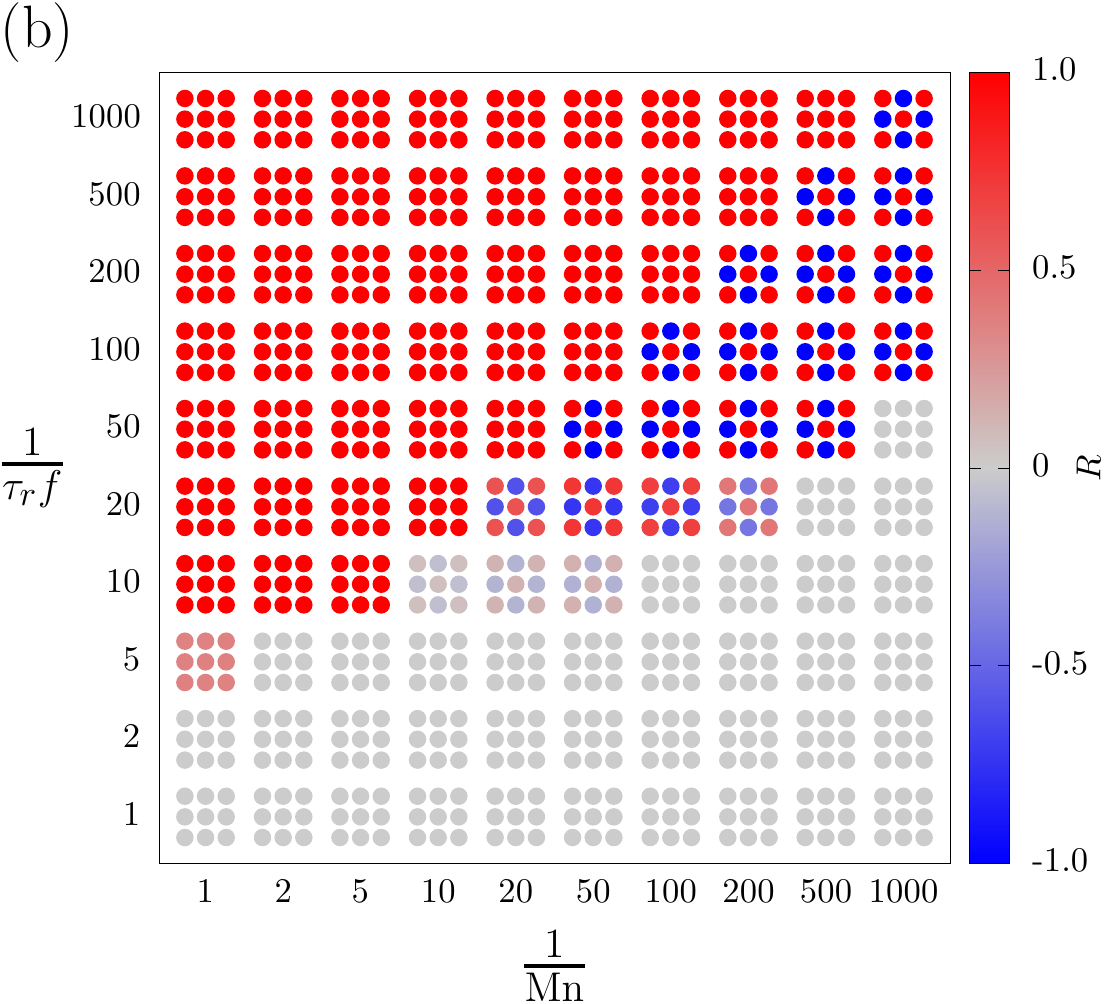}
   \end{center}
 \end{minipage}
 \begin{minipage}{0.42\hsize}
   \begin{center}
   \includegraphics[width=\columnwidth]{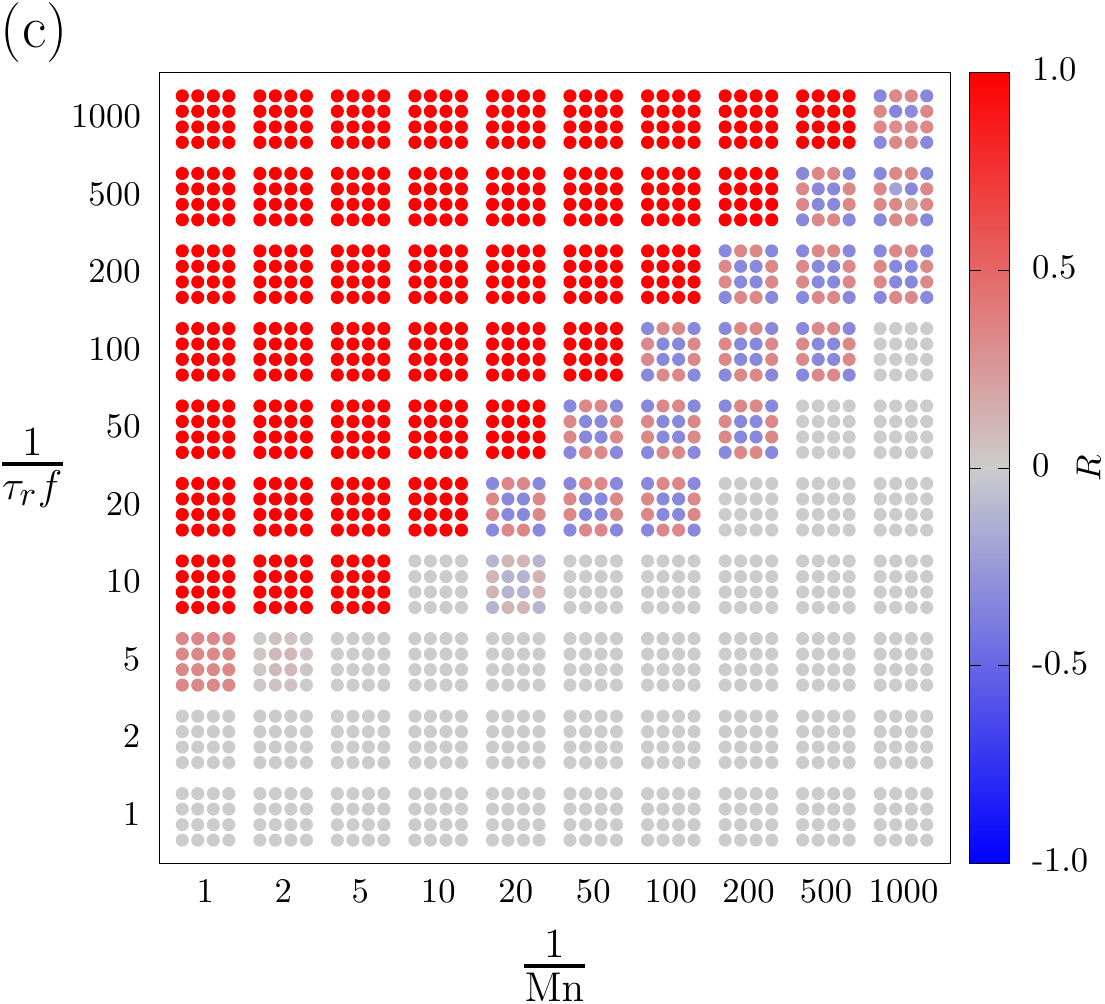}
   \end{center}
 \end{minipage}
   \hspace{0.05\hsize}
 \begin{minipage}{0.42\hsize}
   \begin{center}
   \includegraphics[width=\columnwidth]{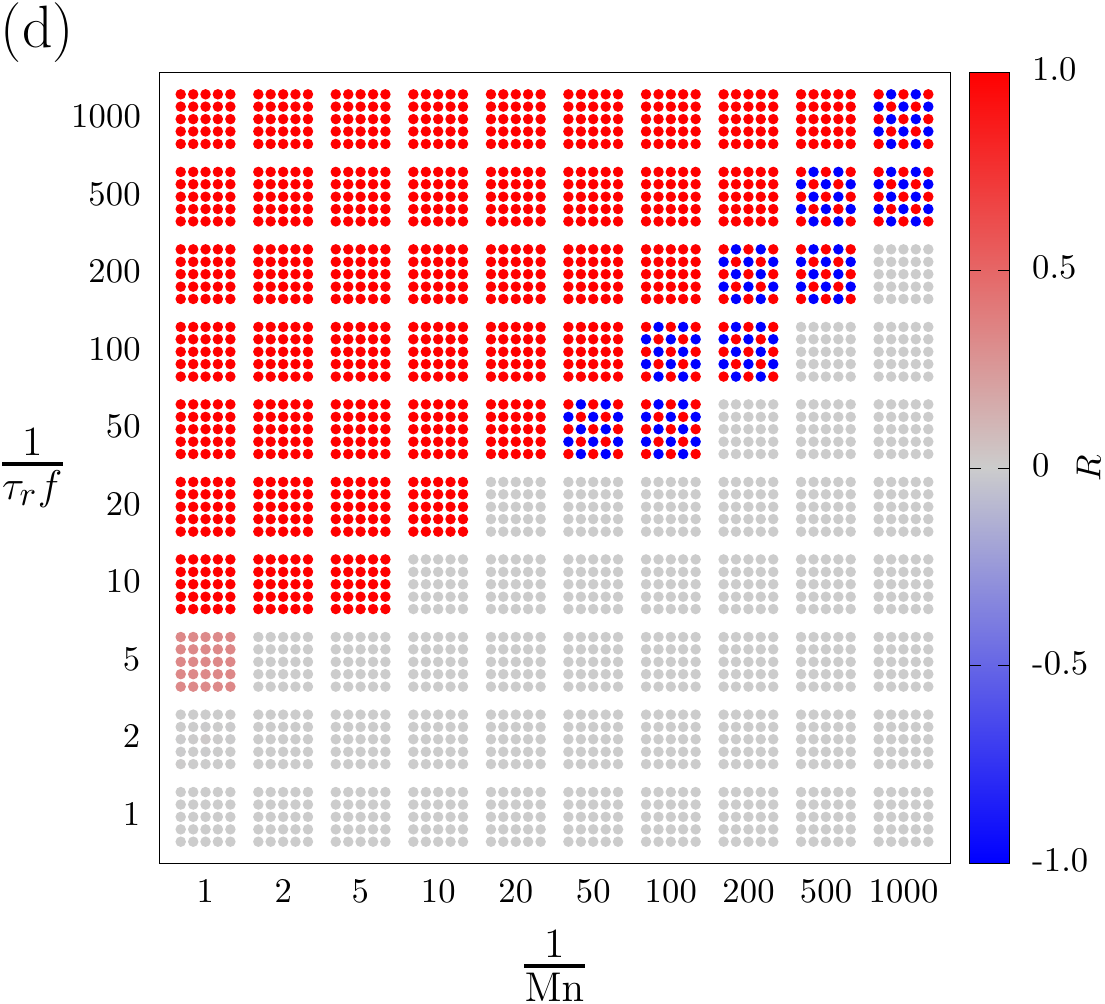}
   \end{center}
 \end{minipage}
 \caption{Phase diagrams for rotor arrays with different sizes: (a) $2\times2$, (b) $3\times3$, (c) $4\times4$, (d) $5\times5$. The color reflects the value of $R$. } \label{fig:arraysize}
\end{figure*}

\subsection{The effect of lattice size}
Figure~\ref{fig:arraysize} shows the phase diagrams for different sizes of the rotor arrays. The value of $R$ as defined in Eq. (\ref{R}) is always in the range $-1 \leq R \leq 1$, and the color in Fig. \ref{fig:arraysize} represents the rotational directions: red ($R = 1$) corresponds to rotation along with the external field, blue ($R = -1$) shows the opposite rotation and gray ($R = 0$) shows rotors with no net rotation.

Although the basic structure of the phase diagram does not depend on the array size, the rotational patterns are different around the regions where the two torque contributions are comparable, namely $1/(\tau_r f) \sim 1/{\rm Mn}$.
The arrays that have odd values of $N_x$ and $N_y$ show the same chessboard pattern as the $3 \times 3$ array (see also \textbf{Movie 3}). By comparing arrays of $3 \times 3$, $5 \times 5$, and $7 \times 7$ (data not shown), we find that the parameter range over which the chessboard pattern is stable becomes smaller with increasing system size.
For $4 \times 4$ arrays, there are equal numbers of clockwise and counterclockwise rotors for $1/(\tau_r f) \sim 1/{\rm Mn}$.  However, the pattern differs from the chessboard pattern: each rotor takes three periods to complete a single full rotation. Moreover, a $2 \times 2$ array shows only synchronized or oscillating patterns.

\subsubsection*{Metachronal waves in a large rotor array}
\begin{figure*}
  \begin{center}
   	\includegraphics[width=0.9\hsize]{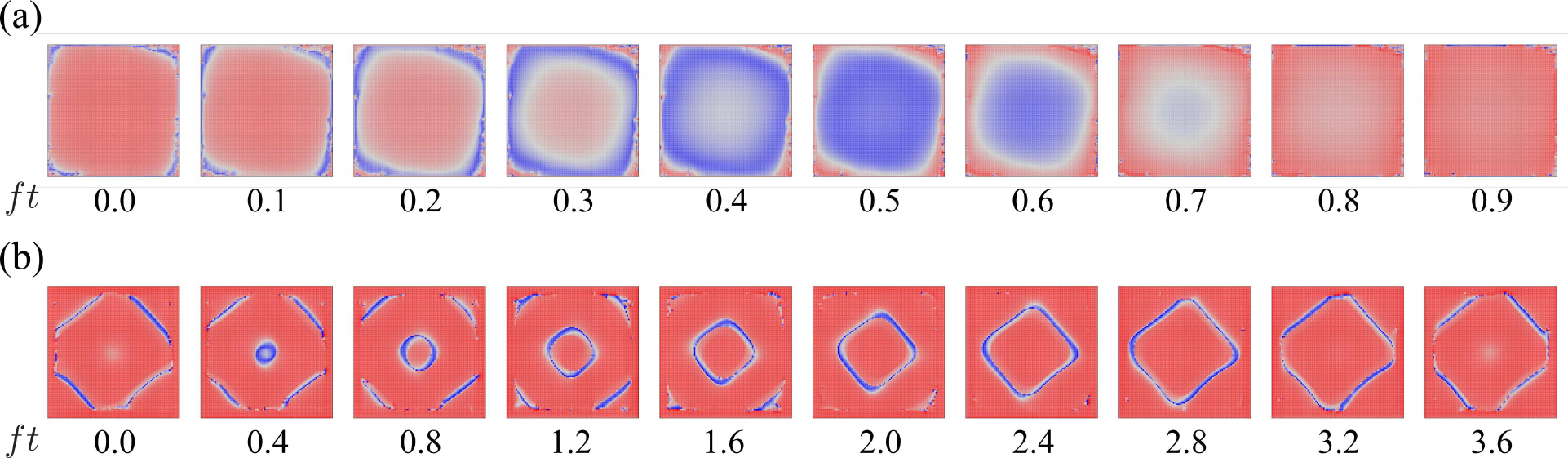}
  \end{center}
  \caption{Propagation of slip waves in a $100 \times 100$ magnet array. (a) Time-series of the rotor velocities from left to right, for $1/\rm{Mn}=1$, $1/\tau_r f=5$ and $a^*=0$. Negative rotational velocity $\omega < 0$ (blue; slipping) propagates from the perimeter to the centre as the time elapses. (b) The slip event propagates from the centre to the perimeter when hydrodynamic interactions between the rotors are switched on: $1/\rm{Mn}=1$, $1/\tau_r f=10$ and $a^*=0.45$.}
\label{fig:slipwave}
\end{figure*}

To examine the effect of the system size even further, we consider the case of $N_x = N_y = 100$. Figure~\ref{fig:slipwave} shows the rotational pattern under a slip condition $1/(\tau_r f) = 5 < 2 \pi$ and $1/{\rm Mn} = 1$. Although the magnets rotate in the same direction as the external field ($\omega > 0$; red) most of the time, they sometimes slip ($\omega < 0$; blue) when they cannot keep up with the field. Interestingly, the slipping motion exhibits a wave-like collective pattern in the large system. We observe that the direction of wave propagation depends on whether or not we incorporate hydrodynamic interactions.

When interactions mediated by the fluid are negligible $a^* = a/\ell \ll 1$, the slipping motion is first triggered at the outer edges of the system and slowly propagates inwards, as shown in Fig.~\ref{fig:slipwave}(a) and \textbf{Movie 4}. This is because the alignment of the outer layers is strongly distorted due to the open boundary. If we now include the first-order hydrodynamic coupling between the rotors (see Appendix \ref{app:fluid}) and increase the effective density by using the parameter $a^* = 0.45$, the direction of the wave propagation reverses and the rotors first slip at the inner layers as shown in Fig.~\ref{fig:slipwave}(b) and \textbf{Movie 5}. This occurs because the hydrodynamic effects are stronger at the centre of the array. As shown by Eq.~(\ref{eq:omega_z}), each rotor creates a flow field with vorticity that is opposite to the rotating direction of the rotor. Therefore the hydrodynamic interactions tend to slow down neighbouring rotors. Since the inner layers have a larger number of neighbours, the slip starts from this region.

\section{Conclusion}
In this paper, we describe the collective motion of magnetic rotors under a rotational magnetic field. We mainly focus on a $3 \times 3$ array of magnets, and observe three characteristic dynamical rotor patterns as the external field strength is varied. In the synchronized regime, which appears when the magnetic torque due to the external field is dominant, all magnets rotate in the same direction as the  field. If, however, the dipolar interaction between the rotors is dominant, all rotors exhibit oscillations (rather than full rotations) during a cycle. When the contributions from the external field and the dipolar interaction are comparable, there is an unexpected chessboard pattern of rotations where the magnets rotate in opposite directions to their closest neighbours. We argue that the chessboard pattern appears as a consequence of the rotor array switching between stripe and spin-ice configurations: essentially the field lifts the degeneracy of these two states. For large system sizes, we observe the propagation of slip waves reminiscent of the metachronal waves observed in natural systems \cite{Uchida2010,Osterman2011,Golestanian2011,Brumley2012,Elgeti2013,Narematsu2015,Brumley2015}.

Such rotor arrays have potential as microfluidic devices that can create various flow fields and vortices of different sizes.
Depending on the external field condition, the flow around the device considered here can vary from a single large vortex, to small alternating vortices around each rotor. We believe that our work sheds light on new collective dynamics of magnetic rotors and opens possibilities for future experiments and applications.

\section*{Acknowledgement}
This work has received funding from the Horizon 2020 research and innovation programme of the EU under Grant Agreement No. 665440, and Osaka University Engineering Science Student Dispatch Program 2018.
F.M. acknowledges the Strategic Priority Research Program of Chinese Academy of Sciences (Grant No. XDA17010504) and the Alexander von Humboldt Foundation for the partial support.

\appendix

\section{Effect of hydrodynamics} \label{app:fluid}
Considering the hydrodynamic coupling between rotors to the leading order \cite{Kim1991,Matsunaga2019}, the angular velocity of the $i$-th rotor is
\begin{equation}
  \omega_i = \frac{d \phi_i}{dt}
  = \frac{T_i}{8 \pi \eta a^3} - \frac{1}{16 \pi \eta} \sum_{j \neq i}^N\frac{T_j}{r_{ij}^3}
  \label{eq:omega_app}
\end{equation}
where the second term in equation~(\ref{eq:omega_app}) is a consequence of the flow field produced by rotation of the other spheres \cite{Kim1991}. The relative strength of the second term is determined by a parameter $a^*=a/\ell$ as is evident from the following dimensionless form:
\begin{eqnarray}
  T_i^* = \frac{1}{\rm Mn}\,T_i^{{\rm dd}*}+\frac{1}{\tau_r f}\,T_i^{{\rm ext}*}, \\
  \omega_i^*=\frac{\omega_i}{f}=T_i^* -\frac{a^{*3}}{2}\sum_{j \ne i}^N\frac{T_j^*}{r_{ij}^{*3}}.
\end{eqnarray}
When the rotor size is sufficiently small, $a \ll \ell$, the hydrodynamic coupling can be neglected, resulting in equation~(\ref{eq:omegaast}).

\section{Favourable configurations} \label{app:B}
Since it is difficult to analyze the energy of a system that has 9 degree of freedoms (the $3 \times 3$ array), we consider a simplified description with a single degree of freedom, the deviation angle $\Delta \phi$, and discuss the resulting energy landscape in this Appendix. We consider the stripe pattern as a reference configuration, with $\Delta \phi = 0$, and define the deviation of the orientation of the $i$-th rotor, located at the $p$-th row and $q$-th column of the array, as
\begin{equation}
  \phi_i = \phi_{p,q}(\Delta\phi) = \pi\cos{(q\pi/2)} + (-1)^{p+q}\Delta\phi. \label{eq:pq}
\end{equation}
The simplified stripe configurations ($\Delta \phi = n \pi/2$) and the simplified spin-ice configurations ($\Delta \phi = (2n + 1)\pi/4$) can thus be described by changing a single parameter $\Delta \phi$.

Varying the deviation $\Delta \phi$ from $0$ to $2\pi$, we find that the magnetic dipole-dipole interaction energy $U^{\rm dd}$ is constant for odd $N' = N_x = N_y$, not only for the stripe and spin-ice configurations, but also for all other angles. On the other hand, the potential $U^{\rm dd}$ for even arrays is not constant. The potentials due to the external magnetic field, $U^{\rm ext}$, are
\begin{eqnarray}
U^{\rm ext*}(\phi_{p,q}(\Delta \phi), \theta_0) =
\left\{\begin{array}{ll}
-N'\cos(\theta_0-\Delta\phi) & (\rm{odd}\;N') \nonumber \\
0 & (\rm{even}\;N') \nonumber \\
\end{array} \right. \\
\end{eqnarray} 
and arrays with even $N'$ show a constant energy level.

As the result, $U^{\rm ext}$ is dominant when $N'$ is odd, and the orientational pattern prefers $\Delta \phi = \theta_0$. For arrays with even $N'$, there is no preferred configuration because $U^{\rm ext}$ is constant. 

\section*{Movie captions}
\begin{itemize}
\item{Movie 1: Typical rotational patterns in a $3\times 3$ array of magnetic rotors: (a) synchronized pattern ($1/{\rm Mn}=100$, $1/(\tau_rf)=500$), (b) chessboard pattern ($1/{\rm Mn} = 100$, $1/(\tau_rf)=50$) and
(c) oscillating pattern ($1/{\rm Mn}=100$, $1/(\tau_rf) = 10)$.
The  arrows at the bottom right indicate the direction of the external magnetic field.
Bold arrows at the beginning of the movie indicate that a strong magnetic field $1/(\tau_r f) = 10^4$ is applied to align the rotors.
}
\item{Movie 2: Three different types of chessboard pattern: (a) chessboard A ($1/{\rm Mn} = 500$, $1/(\tau_rf) = 100$), (b) chessboard A but with $|R|<1$ ($1/{\rm Mn} = 200$, $1/(\tau_rf) = 20$) and (c) chessboard B ($1/{\rm Mn} = 100$, $1/(\tau_rf) = 100$).}

\item{Movie 3: Chessboard pattern in a (a) $5 \times 5$ array and (b) $7 \times 7$ array, for dimensionless parameters $1/{\rm Mn} = 100$ and $1/(\tau_rf) = 100$.}

\item{Movie 4: Slip wave propagation from outer  to inner rotors in a $100 \times 100$ array, for dimensionless parameters $1/{\rm Mn} = 1$, $1/(\tau_rf) = 5$ and $a^*=0$. Note that the rotor size is set as $a^*=0.5$ for the visualization.}

\item{Movie 5: Slip wave propagation from outer to inner rotors in a $100 \times 100$ array, for dimensionless parameters $1/{\rm Mn} = 1$, $1/(\tau_rf) = 10$ and $a^*=0.45$.}
\end{itemize}

\bibliography{reference}

%Create the reference section using BibTeX:

\end{document}